\documentclass[prl,%preprint,tightenlines,%
superscriptaddress,showpacs,floatfix,nofootinbib,twocolumn
]{revtex4}
\usepackage{epsfig}
\newcommand{\slashed}{\not\hspace{-0.7mm}}
\begin{document}
\title{On the definition of the $\Delta$ mass and width}
\author{D.~Djukanovic}
\affiliation{Institut f\"ur Kernphysik, Johannes
Gutenberg-Universit\"at, D-55099 Mainz, Germany}
\author{J.~Gegelia}
\affiliation{Institut f\"ur Kernphysik, Johannes
Gutenberg-Universit\"at,  D-55099 Mainz,
Germany}
\affiliation{
High Energy Physics Institute, Tbilisi State University,
Tbilisi, Georgia}
\author{S.~Scherer}
\affiliation{Institut f\"ur Kernphysik, Johannes
Gutenberg-Universit\"at, D-55099 Mainz, Germany}
\date{\today}

\begin{abstract}
   In the framework of effective field theory we show that, at two-loop
order, the mass and width of the $\Delta$ resonance defined via the
(relativistic) Breit-Wigner parametrization both depend on the choice of field
variables.
   In contrast, the complex-valued position of the pole of the propagator is
independent of this choice.

\end{abstract}

\maketitle

   The problem of defining masses of unstable particles has a long history.
   A popular definition corresponding to a (relativistic) Breit-Wigner formula
makes use of the zero of the real part of the inverse propagator to identify the
mass.
   The field-redefinition dependence of such a definition was shown
in Refs.~\cite{Willenbrock:1990et,Valencia:1990jp} in the scalar sector of the
Standard Model.
   Another important example %in the context of the Standard Model
is the definition of the $Z$-boson mass.
   The gauge-parameter dependence of the Breit-Wigner mass
starting at two-loop order was shown in
Refs.~\cite{Sirlin:1991fd,Sirlin:1991rt,Willenbrock:1991hu,Gegelia:1992kj,Gambino:1999ai}.
   In contrast, defining the mass and width in terms of the complex-valued
position of the pole of the propagator leads to both field-redefinition and
gauge-parameter independence \cite{'tHooft:1972ue,Lee:1972fj,Balian:1976vq}.

It was noted in Ref.~\cite{Willenbrock:1991hu} that, as there is no fundamental
theory of baryon resonances, the issue of field-redefinition invariance and
gauge-parameter (in)dependence does not arise for unstable particles of this
kind. As these resonances are thought to be described by QCD, with the progress
of lattice techniques and, especially, the low-energy effective theories (EFT) of
QCD (see,
e.g., \cite{Weinberg:1979kz,Gasser:1983yg,Gasser:1988rb,Scherer:2005ri,%
Hemmert:1997ye,Hacker:2005fh,Pascalutsa:2006up} and references therein) the
question of defining resonance masses becomes important.
   In this letter we
examine this issue for the $\Delta$ resonance.
   As discussed in Ref.~\cite{Hoehler}, the {\it
conventional resonance mass} and width determined from generalized Breit-Wigner
formulas have problems regarding their relation to S-matrix theory and suffer
from a strong model dependence.
   Here, we will show that these parameters, in addition, depend on the field-redefinition
parameter in a low-energy EFT of QCD.

   For simplicity we ignore isospin and consider an EFT of a single
nucleon, pion, and $\Delta$ resonance.
   Defining
$$
\Lambda_{\mu\nu}=-(i\slashed{\partial}-m_\Delta) g_{\mu\nu}+i
\,(\gamma_{\mu}\partial_{\nu}+\gamma_{\nu}\partial_{\mu}) - i
\gamma_{\mu}\slashed{\partial}\gamma_{\nu} -m_\Delta
\gamma_{\mu}\gamma_{\nu}, %\label{lambdaA} \nonumber
$$
   the free Lagrangian is given by
\begin{equation}
\mathcal{L}_{0}=
\bar{\psi}^{\mu}\,\Lambda_{\mu\nu}\,\psi^{\nu}+\bar\Psi(i\slashed{\partial}-m_N)\Psi
+\frac{1}{2}\,\partial_\mu\pi \partial^\mu\pi \,. \label{LfreiA}
\end{equation}
   Here, the vector-spinor $\psi^{\mu}$  describes
the $\Delta$ in the Rarita-Schwinger formalism \cite{Rarita:1941mf},
   $\Psi$ stands for the nucleon field with mass $m_N$, and $\pi$
represents the pion field which we take massless to simplify the calculations.
   The most general (free) Rarita-Schwinger Lagrangian contains an arbitrary parameter $A$.
   In a consistent theory having the right number of degrees of freedom,
physical observables do not depend on $A$ \cite{Wies:2006rv} and we
have chosen a convenient value, namely $A=-1$. We consider the
interaction terms of the form
\begin{equation}
\mathcal{L}_{\rm int}= g\, \partial^\nu \pi\, \bar\psi^\mu \left(
g_{\mu\nu}- z \gamma_\mu\gamma_\nu \right)\Psi +
\mbox{H.c.}+\cdots\,, \label{Intlagrangian}
\end{equation}
where the ellipsis refers to an infinite number of interaction terms
which are present in the EFT. These terms also include all
counter-terms which take care of divergences appearing in our
calculations.  The consistency of the interaction terms with the
constraints of the spin-3/2 system fixes the value of the parameter
$z$ to $-1$ for $A=-1$ \cite{Nath:1971wp,Hacker:2005fh}. Throughout
this paper we use dimensional regularization. Although our results
are renormalization scheme independent, for simplicity we use the
minimal subtraction scheme \cite{cbare}. It is implemented by
subtracting the divergent parts of one- and two-loop diagrams using
the standard procedure \cite{Collins:1984xc}.

Let us consider the field transformation
\begin{equation}
\bar\psi^\mu\to \bar\psi^\mu + \xi\,\partial^\mu\pi \bar\Psi\,,\quad \psi^\nu\to
\psi^\nu+\xi \,\partial^\nu\pi \Psi\,,  \label{ftr}
\end{equation}
where $\xi$ is an arbitrary real parameter.
   When inserted into the Lagrangians of Eqs.~(\ref{LfreiA}) and
(\ref{Intlagrangian}), the field redefinition generates additional interaction
terms.
   The terms linear in $\xi$ are given by
\begin{equation}
\mathcal{L}_{\rm add\, int}= \xi\,\partial^\mu\pi\,
\bar\Psi\,\Lambda_{\mu\nu}\,\psi^{\nu}+ \xi \,
\partial^\nu\pi\, \bar{\psi}^{\mu}\,\Lambda_{\mu\nu}\,\Psi %- \xi \,
%\partial^\beta\partial^\mu \pi\, \bar{\psi}^{\alpha}\,\Lambda_{\alpha\beta\mu}\,\Psi\,,
\,. \label{NewIntTerms}
\end{equation}
   Note that the contribution generated from the expression explicitly shown in
Eq.~(\ref{Intlagrangian}) vanishes identically.
   Because of the equivalence theorem
physical quantities cannot depend on the field redefinition parameter $\xi$.
   Below we demonstrate that the complex-valued position of the pole of the $\Delta$
propagator does not depend on $\xi$. In contrast, the mass and width
defined via (the zero of) the real and imaginary parts of the
inverse propagator depend on $\xi$ at two-loop order.

The dressed propagator of the $\Delta$ in $n$ space-time dimensions
can be written as \cite{Hacker:2005fh,note}
\begin{eqnarray}
&-&i \left[ g^{\mu\nu}-\frac{\gamma^\mu\gamma^\nu}{n-1} -\frac{
p^\mu\gamma^\nu-\gamma^\mu p^\nu}{(n-1) m_{\Delta}}-\frac{(n-2)
p^\mu p^\nu }{(n-1) m_{\Delta}^2}\right]\nonumber\\
&& \times\frac{1}{p\hspace{-.35 em}/\hspace{.1em}-m_\Delta -\Sigma_1 -
p\hspace{-.4 em}/\hspace{.1em}\Sigma_6} + \mbox{pole-free
terms}\,,\label{dressedDpr}
\end{eqnarray}
where we parameterized the self-energy of the $\Delta$ as
\begin{eqnarray}
&& \Sigma_1(p^2) g^{\mu\nu}+\Sigma_2(p^2)
\gamma^{\mu}\gamma^{\nu}+\Sigma_3(p^2) p^{\mu}\gamma^{\nu}
+\Sigma_4(p^2) \gamma^{\mu}p^{\nu}\nonumber\\
&&+\Sigma_5(p^2)\,p^{\mu}p^{\nu} + \Sigma_6(p^2)\,
p\hspace{-.45em}/\hspace{.1em}
g^{\mu\nu}+\Sigma_7(p^2)\,p\hspace{-.45em}/\hspace{.1em}
\gamma^{\mu}\gamma^{\nu}\nonumber\\
&& +\Sigma_8(p^2)\, p\hspace{-.45em}/\hspace{.1em}
p^{\mu}\gamma^{\nu}+\Sigma_9(p^2)\, p\hspace{-.45em}/\hspace{.1em}
\gamma^{\mu}p^{\nu}+\Sigma_{10}(p^2)\,p\hspace{-.45em}/\hspace{.1em}
p^\mu p^\nu. \label{DseParametrization}
\end{eqnarray}
   The complex pole $z$ of the $\Delta$ propagator is obtained by solving the
equation
\begin{equation}
z - m_\Delta -\Sigma_1(z^2)-z\, \Sigma_6(z^2)=0\,.
\label{poleequation}
\end{equation}
The pole mass is defined as the real part of $z$.

On the other hand, the mass $m_R$ and width $\Gamma$ of the $\Delta$ resonance
are often determined from the real and imaginary parts of the inverse propagator
(corresponding to the Breit-Wigner parametrization), i.e.,
\begin{eqnarray}
&& m_R - m_\Delta -{\rm Re}\, \Sigma_1(m_R^2)-m_R \, {\rm Re}\,
\Sigma_6(m_R^2)=0\,,\nonumber\\
&&\Gamma = -2\, {\rm Im}\, \Sigma_1(m_R^2)-2\, m_R \, {\rm Im}\,
\Sigma_6(m_R^2) \,. \label{Rmassequation}
\end{eqnarray}
   Below we calculate the $\Delta$ mass using both definitions and analyze their
$\xi$ dependence to first order.

\begin{figure}
\epsfig{file=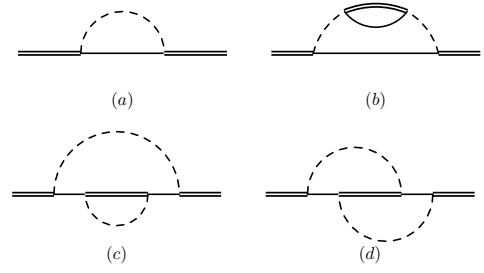,width=0.35\textwidth}
\caption[]{\label{DeltaMassInd:fig} $\Delta$ self-energy diagrams.
Solid, dashed, and double lines correspond to nucleon, pion, and $\Delta$,
respectively.}
\end{figure}

The $\Delta$ self-energy at one loop-order is given by the diagram
in Fig.~\ref{DeltaMassInd:fig} (a). The corresponding results for
$\Sigma_1$ and $\Sigma_6$ read
\begin{eqnarray}
\Sigma_1^{(a)}&=&-g^2\,m_N\,I_1 - 2 \xi\, g
\left[\left(p^2-m_\Delta m_N\right) I_1  + p^2 J_1\right], \nonumber\\
\Sigma_6^{(a)}&=&-g^2 \left(I_1 + J_1\right) + 2 \xi\, g \left[
m_\Delta J_1 + \left(
m_\Delta-m_N\right) I_1 \right]\,,\nonumber\\
\label{DsePar}
\end{eqnarray}
where $I_1$, $J_1$ are defined through the one-loop integrals
\cite{note2}
\begin{eqnarray}
I^{\alpha\beta}\,, I^{\alpha\beta\gamma} & = & \int \frac{i\, d^nk}{(2\,\pi)^n}
\frac{k^\alpha k^\beta, k^\alpha k^\beta k^\gamma }{\left[
k^2+i\,0^+\right]\left[ \left(p+k\right)^2-m_N^2+i\,0^+\right]},\nonumber
\label{integrals}
\end{eqnarray}
which we parameterize as
\begin{eqnarray}
I^{\alpha\beta} & = & I_1 g^{\alpha\beta}+I_2 p^\alpha
p^\beta\,,\nonumber\\
I^{\alpha\beta\gamma} & = & J_1\left(g^{\alpha\beta}p^\gamma
+g^{\alpha\gamma}p^\beta + g^{\beta\gamma}p^\alpha \right)+J_2 p^\alpha p^\beta
p^\gamma . \nonumber \label{intpar}
\end{eqnarray}

The two-loop contributions to the $\Delta$ self-energy are given in
Fig.~\ref{DeltaMassInd:fig} (b) - (d). We are interested in terms
linear in $\xi$. Calculating diagram (b) and (c) we find that they
give vanishing contributions.  The result of diagram (d), linear in
$\xi$, can be reduced to the form
\begin{eqnarray}
\Sigma_1^{(d)} & = & 2\,g^3 \xi\,\left[ m_N^2
I_1^2 + p^2 \left(I_1 + J_1\right)^2 \right]\,, \nonumber\\
\Sigma_6^{(d)} & = & 4\,g^3\,\xi\,m_N\,I_1 \left(I_1 + J_1\right)\,.
\label{Dse2loop}
\end{eqnarray}

Note that the vanishing of the contributions of diagrams (b) and (c)
as well as the simple expression of Eq.~(\ref{Dse2loop}) have to be
attributed to the special choice of the field transformation of
Eq.~(\ref{ftr}).

To find the pole of the propagator we insert its loop expansion
\begin{equation}
z=m_\Delta+\delta_1 z+\delta_2 z+\cdots\, \label{polesparametr}
\end{equation}
in Eq.~(\ref{poleequation}) and solve the resulting equation order by order.
  Using Eq.~(\ref{DsePar}) we obtain for the one-loop result
\begin{eqnarray}
\delta z_1 & = & -g^2\,\left[ m_\Delta \bar J_1 + \left(
m_\Delta+m_N\right) \bar I_1
\right]\,, \nonumber\\
&& \bar J_1=J_1|_{p^2=m_\Delta^2}\,, \ \ \ \bar
I_1=I_1|_{p^2=m_\Delta^2}. \label{deltam1}
\end{eqnarray}
   The contribution to the two-loop expression $\delta z_2$, linear in $\xi$,
generated by the one-loop diagram reads
\begin{equation}
\delta z_{2,1 L}^\xi= 2\,g^3\,\xi\,\left[ m_\Delta \bar J_1 +\left(
m_\Delta+m_N\right) \bar I_1 \right]^2\,. \label{deltam2}
\end{equation}
For the genuine two-loop contribution to $\delta z_2$ we have
\begin{equation}
\delta z_{2,2 L}^\xi=-2\,g^3\,\xi\,\left[ m_\Delta \bar J_1 + \left(
m_\Delta+m_N\right) \bar I_1 \right]^2. \label{deltam22loop}
\end{equation}
These two contributions exactly cancel each other leading to the
$\xi$-independent pole of the propagator.

We perform the same analysis inserting the loop expansion of $m_R$,
\begin{equation}
m_R=m_\Delta+\delta_1 m+\delta_2 m + \cdots,  \label{Rmassparametr}
\end{equation}
in Eq.~(\ref{Rmassequation}). For $\delta m_1$ we obtain
\begin{equation}
\delta m_1=-g^2\,\left[ m_\Delta {\rm Re}\, \bar J_1 +\left(
m_\Delta+m_N\right) {\rm Re}\, \bar I_1 \right]\,.
 \label{deltam1R}
\end{equation}
The contribution to $\delta m_2$ generated by the one-loop diagram reads
\begin{equation}
\delta m_{2,1 L}^\xi= 2\,g^3\,\xi\,\left[ m_\Delta {\rm Re}\,\bar
J_1 + \left( m_\Delta+m_N\right) {\rm Re}\, \bar I_1 \right]^2\,.
\label{deltam2R}
\end{equation}
For the two-loop contribution to $\delta m_2$ we have
\begin{equation}
\delta m_{2,2 L}^\xi=-2\,g^3\,\xi\,{\rm Re} \left[ m_\Delta \bar J_1
+ \left( m_\Delta+m_N\right) \bar I_1 \right]^2.
\label{deltam22loopR}
\end{equation}
For an unstable $\Delta$ resonance $\bar I_1$ and $\bar J_1$ have imaginary parts
and therefore Eqs.~(\ref{deltam2R}) and (\ref{deltam22loopR}) do {\em not} cancel
each other, thus leading to a $\xi$-dependent mass $m_R$. An analogous result
holds for the width $\Gamma$ obtained from Eq.~(\ref{Rmassequation}).

\medskip

To conclude, we addressed the issue of defining the mass and width
of the $\Delta$ resonance in the framework of a low-energy EFT of
QCD. In general, the scattering amplitude of a resonant channel can
be presented as a sum of the resonant contribution expressed in
terms of the dressed propagator of the resonance and the background
contribution. The resonant contribution defines the Breit-Wigner
parameters through the real and imaginary parts of the inverse
(dressed) propagator. The resonant part and the background
separately depend on the chosen field variables, while the sum is of
course independent of this choice. We have performed a particular
field transformation with an arbitrary parameter $\xi$ in the
effective Lagrangian. In a two-loop calculation we have demonstrated
that the mass and width of the $\Delta$ resonance determined from
the real and imaginary parts of the inverse propagator depend on the
choice of field variables. On the other hand, the complex pole of
the full propagator does not depend on the field transformation
parameter $\xi$.

Note that according to general theorems
\cite{Chisholm:1961,Kamefuchi:1961sb,Coleman:1969sm,Weinberg:1995mt}
it is expected that in quantum field theories the $S$-matrix is
independent of field redefinitions (change of variables). The pole
of the $\Delta$ propagator corresponds to the pole of the
$S$-matrix. The Laurent-series expansion of the $S$-matrix around
this pole has to be independent of the field redefinition
term-by-term. Therefore the pole of the $\Delta$ propagator is
expected to be independent of the field redefinition.

The conclusions from this work are not restricted to the considered
toy model or EFT in general. Rather, our results are a manifestation
of the general feature that the (relativistic) Breit-Wigner
parametrization leads to model- and process-dependent masses and
widths of resonances. Although in some cases (like the $\Delta$
resonance) the background has small numerical effect on the
Breit-Wigner mass, still the pole mass and the width should be
considered preferable as these are free of conceptual ambiguities.
This agrees with and supports the recent results of
Ref.~\cite{Bohm:2004zi}.

\acknowledgments

We would like to thank L.Tiator for useful discussions. D.D. and
J.G. acknowledge the support of the Deutsche Forschungsgemeinschaft
(SFB 443).


\begin{thebibliography}{99}
\bibitem{cbare} For the purposes of this work we could have also
worked in terms of bare parameters. In this case no counter term
diagrams occur and the loop integrals do not need to be subtracted.
\bibitem{note2} The loop integrals $I_1$ and $J_1$ are made finite
by applying the minimal subtraction scheme, i.e., by subtracting the
parts proportional to $1/(n-4)$.
\end{thebibliography}
\end{document}